# Planck's uncertainty principle and the saturation of Lorentz boosts by Planckian black holes [#]

A. Aurilia [*]

Department of Physics

California State Polytechnic University

Pomona, CA 91768

E. Spallucci[**]

Dipartimento di Fisica Teorica

Universita` di Trieste

INFN, Sezione di Trieste

**Abstract**

A basic inconsistency arises when the *Theory of Special Relativity* meets with quantum phenomena at the Planck scale. Specifically, the Planck length is Lorentz invariant and should not be affected by a Lorentz boost. We argue that Planckian relativity must necessarily involve the effects of black holes formation. Recent proposals for resolving the noted inconsistency seem unsatisfactory in that they ignore the crucial role of gravity in the saturation of Lorentz boosts. Furthermore, an invariant length at the Planck scale amounts to a universal quantum of resolution in the fabric of spacetime. We argue, therefore, that the universal Planck length requires an extension of the Uncertainty Principle as well. Thus, the noted inconsistency lies at the core of Quantum Gravity. In this essay we reflect on a possible resolution of these outstanding problems.

[*] E-mail : aaurilia@csupomona.edu
[**] E-mail : spallucci@trieste.infn.it

## Introduction

There is a longstanding conceptual problem in high-energy physics that only recently has emerged as a compelling question in need of a definite answer.

In its simplest and most irreducible form, the problem is that of reconciling Planck scale discreteness of quantum spacetime with the Lorentz-Fitzgerald contraction of lengths.

It is often argued that in a quantum theory of gravity the Planck length plays the role of minimal universal length [1]. However, the concept of a universal extremal length in Nature seems to be in conflict with the requirement of Lorentz invariance. Indeed, since lengths transform continuously under a Lorentz transformation, the Planck length ought to contract in a Lorentz boost. Then, in a suitable reference frame, an observer ought to see a *contracted* Planck length, shorter than the original *frame independent* length assumed to be minimal. This is the contradiction to be resolved.

The existing literature does not reflect the fact that a possible resolution of the apparent conflict between Lorentz transformations and Planck scale discreteness involves ***two*** fundamental modifications, one in relativity, the other in quantum mechanics, the two cornerstones of contemporary physics. The first necessary modification is the saturation of Lorentz boosts at the Planck scale, already noted in the literature [2], the second is an equally necessary extension of Heisenberg's uncertainty principle at the same scale. A generalized form of the uncertainty relation has been proposed in the framework of String Theory and in a model independent way by M. Maggiore [3], but never connected to the saturation of Lorentz boosts. However, it seems to us that the two modifications go hand in hand in the sense that accepting the existence of a minimal universal length is tantamount to introducing a quantum of resolution, and therefore uncertainty, in the measurement of distances.

To date, two types of extension of special relativity have been suggested that incorporate two invariant scales, the speed of light and the Planck length, or energy. The first [4], somewhat improperly dubbed "Doubly Special Relativity," belongs to a class of models derived from the kappa-Poincare' algebra [5], a deformation of the Poincare` algebra, while the second corresponds to a non linear realization of the boost operator of the usual Lorentz group [6].

A deformation of the Poincare` algebra may well turn out to be the correct framework not only for an extension of relativity at the Planck scale but also for an extension of the uncertainty relation [7] and the idea of non commutativity of quantum spacetime [8].[1] However, there is an infinity of choice of bases of the kappa-Poincare` algebra and a physical criterion for selecting one without ambiguity is still lacking. Be that as it may, *a drawback of the existing models of Planckian relativity [4-6], is that there is no mention of black hole formation, a natural occurrence in a Planckian environment, nor is there any mention of the second modification required by the new relativity, namely the extension of the quantum Uncertainty Principle.*
In the following, arguments are presented that take into account the above requirements.

**A dimensional argument for a minimal universal length**

Planck's system of units is built out of three fundamental physical constants: c, h and G. As is well known, the Planckian standards of length, mass and time are expressed, in conventional units and *up to a numerical normalization factor*, as follows,

$$L_P \propto \sqrt{hG/c^3} \; ; \qquad M_P \propto \sqrt{ch/G} \; ; \qquad T_P \propto \sqrt{hG/c^5}.$$

As one readily verifies, in terms of c, h and G, *and again up to a numerical normalization factor,* to any mass M one can associate two physical lengths, namely, the Compton wavelength, $L_C \propto \frac{h}{Mc}$ and the Schwarzschild radius, $L_S \propto \frac{GM}{c^2}$.

It is an obvious, but important observation that the Compton wavelength, inversely proportional to mass, does not depend on the gravitational constant G, whereas the Schwarzschild radius, proportional to mass, does not depend on the quantum of action h. This special dependence of the two lengths on the fundamental constants is, of course, a reflection of the framework in which either length makes physical sense: particle physics without gravity for the Compton wavelength, gravity without quantum mechanics for the Schwarzschild radius. However, if $L_C$ is set equal to $L_S$, then the mass M must be proportional to the Planck mass as given above, and one is in the realm of Quantum Gravity.

---

[1] It seems remarkable that many of the contemporary ideas about the short distance structure of spacetime were anticipated in a seminal paper by H. Snyder, see Ref.[1].

Thus, the Planck mass is the unique mass for which the Compton wavelength is of the same order of magnitude as the Schwarzschild radius.

A "Planckion," in this context, is a particle whose mass is of the order of the Planck mass[2]. Since the Schwarzschild radius defines the *event horizon* of a physical object, any particle whose Compton wavelength is less or equal than its Schwarzschild radius is a *black hole*. Therefore, a Planckion is at the physical threshold of being, or becoming a black hole, and its interior, or structure, is hidden from the physically accessible universe.

The implication of a Planckion state as an extremal state of mass-energy, is visually displayed in the (L, m) diagram (sketchy and not to scale) below:

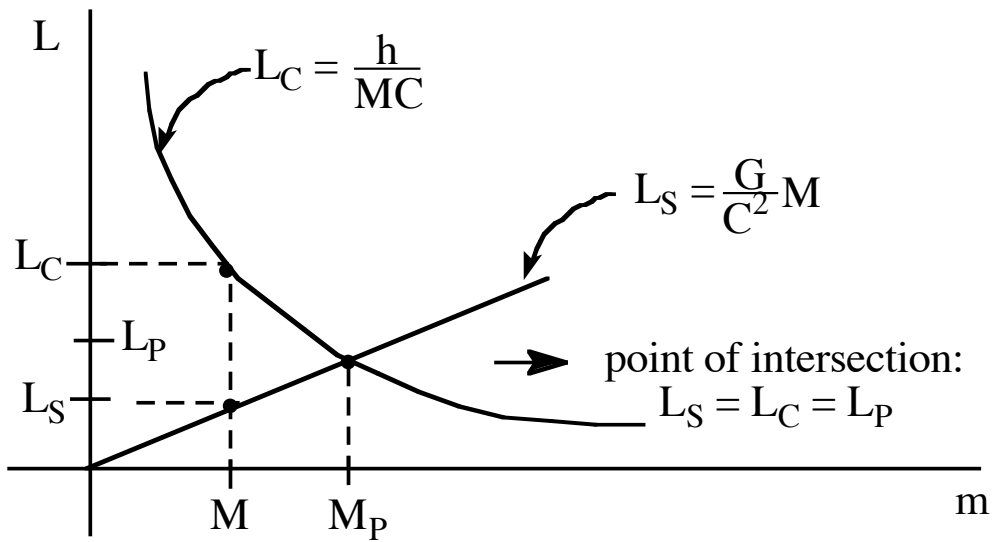

As $M \Rightarrow M_P$, as it might happen in a hypothetical particle accelerator, both $L_C$ and $L_S \Rightarrow L_P$. From here one may be tempted, not unreasonably in the opinion of the authors, to interpret $M_P$ as an extremal value, an upper limit in fact, in the spectrum of physical masses attainable by "elementary particles" to the extent that an elementary particle is characterized by its Compton wavelength, a long standing criterion in Particle Physics. A hypothetical *elementary* particle more massive than a Planckion would have $L_C \prec L_P \prec L_S$, which is the signature of a black hole for which the Compton wavelength no longer represents a physical attribute but is replaced by the Schwarzschild radius of a macroscopic test body. At this stage, the diagram merely reflects the noted dependence of the two characteristic lengths on mass and that the Planck length represents the characteristic scale that divides microscopic objects, for which gravitational effects are negligible, from macroscopoic ones for which gravity is a predominant force. However, the ( L, m) diagram can be given a deeper quantum mechanical interpretation: *for values of L and M near the Planck scale it may be interpreted as a graphical representation of Planck's Uncertainty Principle, a natural extension of Heisenberg's relation at the Planck scale*

$$\Delta x \geq h/\Delta p + (L_P^2/h) \Delta p$$

once we identify $\Delta x \approx L$ and $\Delta p \approx Mc$.[3] In making this identification one must keep in mind that a Planckion is a quantum object by definition, since $L_P$ contains the quantum of action among its constituent fundamental constants. Thus a Planckion, an object the size of $L_P$ is necessarily *fuzzy* to the extent that quantum fluctuations at that scale are of the same order of magnitude as $L_P$. According to this quantum interpretation, the linearly rising branch of the ( L, m) diagram corresponds to the second term in brackets in Planck's uncertainty relation. The standard procedure, at this point, is to minimize the position uncertainty leading to the fundamental identification $(\Delta x)_{min} \propto L_{pl}$, the unique value that saturates Planck's quantum inequality.

---

[2] It seems that the term 'Planckion' as well as 'Maximon' were used in the Russian literature. We are borrowing it from a paper by H. Treder [2].

---

[3] The form of this inequality is identical to the 'string uncertainty relation' except that the value of the constant ( $L_P^2/h$, up to a numerical constant, in our case) is proportional to the string tension $\alpha'$. The remarkably simple extrapolation of the uncertainty relation at the Planck scale from the ( L, m ) diagram in the text is a shortcut of a far more technical derivation recently proposed by the authors in the framework of the theory of p-branes [9].

Thus, the size of the horizon of a black hole is sharply defined by its Schwarzschild radius in classical physics but becomes uncertain in quantum gravity and *cannot be resolved on a scale smaller than* $L_P$. This conclusion agrees with the analysis of a 'gedanken' experiment for the measurement of the area of the horizon of a black hole in quantum gravity [3].

**Saturation of Lorentz boosts by Planckian black holes**

Planck's uncertainty relation implies a *quantum constraint* in the physical interpretation of the ( L, m ) diagram. The unequivocal message of the previous discussion is that $L_C$ and $L_S$, classically unbound, represent mutually *exclusive* characteristic lengths of an object of mass M and can coexist only for a Planckian black hole, in which case both coalesce into a *minimal and universal quantum of resolution,* namely the Planck length. A corollary of this statement, probably not widely appreciated, is that *classically one may **formally** assign a Schwarzschild radius to an elementary particle just as one may formally assign a Compton wavelength to a black hole. However, if one believes in the validity of Planck's uncertainty relation, these assignments make no physical sense since in either case one is led to contemplate a resolution that is smaller than the Planck length*. This consideration, based on the unconventional Planck's uncertainty relation, is one of two key points leading to the saturation of Lorentz boosts and to the preservation of Lorentz invariance at Planckian energies. Our second basic consideration is equally unconventional and stems again from the ( L, m ) diagram.

A superficial glance at that diagram might lead one to the erroneous anticipation that it is the linear dependence of the Schwarzschild radius on mass-energy that provides the **turning point** in the contraction of lengths at the Planck scale. Thus, we hasten to emphasize that the increasing mass-energy envisaged in the ( L, m ) diagram above is not directly related to the Lorentz boosts envisaged in the theory of special relativity. A Lorentz boost (and the concomitant effects of mass increase and length contraction) is a kinematical effect, symmetric between two inertial observers, due to a *coordinate transformation* between two reference frames in relative motion with a pre-assigned velocity bounded only by the speed of light. Pedantic as this may sound, it points to an essential, but largely unnoticed difference between $L_C$ and $L_S$ under a Lorentz boost: while $L_C$ is consistent with the phenomena of mass increase and length contraction, ($L_C$ is inversely proportional to mass) $L_S$ is not, precisely because of its linear dependence on M, so that boosting the mass results in an increase in length. This paradox arises because of a misinterpretation, or rather, a *special relativistic* interpretation of $L_S$. *The way out of this paradox is that $L_S$ is, in fact, invariant under a coordinate transformation of the Schwarzschild metric*. The point of fact is that a naïve application of the Lorentz contraction formula to $L_S$ is fundamentally flawed in that it ignores the necessary effect of gravity, namely the effect of the spacetime geometry that is not Minkowskian. In other words, one must keep in mind that the object of a boost in the case of $L_S$ is the whole Schwarzschild geometry and this involves the metric on the event horizon. Thus, in order to make our point, we shall omit a discussion of the general case and suppose that a Planckion state is realized in Nature by whatever means one dares to conceive.[4] As argued earlier, such an extreme state of mass-energy is necessarily a black hole. Once a Planckian black hole is formed, there is nothing, in principle, that prevents a kinematical boost of the Schwarzschild **coordinates** to any other inertial frame of reference. Under such a boost one may expect a deformation of the event horizon of the Planckian black hole due, for instance, to a contraction of its characteristic length, namely, the Planck length, in the direction of the boost with a concomitant increase in mass energy. This would amount to the basic inconsistency noted in the introduction, namely, a violation of Lorentz invariance accompanied by the possibility of an elementary particle with a mass-energy larger than the Planck mass. However, this is not the case. We have reanalyzed the effect of a boost on the Schwarzschild metric induced by a massive body.[5] *Our analysis is based on two basic properties of the Schwarzschild geometry, namely, a) the area of the event horizon can never decrease, and b) the event horizon is a **null surface** whose shape and size cannot be altered by a change of coordinates*. Our calculations are too long and involved to be reported here. However, they confirm these expectations. A Lorentz boost results merely in a *rearrangement* of the Schwarzschild geometry in terms of the boosted coordinates, leaving unchanged its characteristic length, namely the Schwarzschild radius. In general, the event horizon of *any* black hole is *invariant* under Lorentz transformations of its geometry. A Planckion is no exception but for the fact that its *Schwarzschild radius saturates Planck's quantum inequality.* To reiterate the basic point of this discussion: the Planck length, built out of c, h and G, three universal physical constants, is itself just as universal as c, h and G, and cannot be contracted without violating Lorentz invariance. Hence the motivation, in the current literature, for the search of a " Special Relativity with Two Invariant Scales." To our mind, however, the current models miss the natural solution provided by gravity in the form of a boosted Schwarzschild geometry. *Indeed, it is the necessary onset of gravity at the Planck scale, with its stringent* **classical and quantum limits** *on the area of a black hole, that provides a natural* **end point**, *not a turning point, in the relativistic contraction of any length below the* **Schwarzschild radius of a Planckion**, *namely the Planck length.*

---

[4] The possibility of lowering the Planck threshold within the reach of the next generation of accelerators, as suggested in some theories of gravity in higher dimensions is not contemplated here.

[5] The effect of a boost on the Schwarzschild metric was studied long ago [10], but in the mathematically cumbersome limit of *vanishing rest mass*. In that limit, the apparent effect of a boost is to induce a gravitational shock wave, a region of singularity in an otherwise flat manifold. However, the whole mathematical procedure is ambiguous, if not flawed, in the opinion of the authors, since it represents an attempt to artificially circumvent the fact that a photon cannot be boosted anywhere.

In conclusion, our analysis supports the hypothesis that no physical test body can probe distances shorter than $L_{planck}$, a lower bound that saturates Planck's quantum inequality. Furthermore, this hypothesis is consistent with Lorentz invariance as long as one recognizes the pivotal role of Planckian black holes in the saturation of Lorentz boosts *through the invariance of the event horizon of the Schwarzschild geometry*. It is somewhat satisfying that all of the above is encapsulated in the simple diagram introduced at the beginning of this essay.